\documentclass{article}
\usepackage{pifont}

\usepackage[utf8]{inputenc}
\usepackage[margin=0.75in]{geometry}
\usepackage{import}
\usepackage{caption}
\usepackage{amsmath}
\usepackage{graphicx}
\usepackage{caption}
\usepackage[colorlinks=true, allcolors=blue]{hyperref}
\usepackage{natbib}
\usepackage{booktabs}
\usepackage{tabularx}
\usepackage{float}

\title{Revisiting the Resource Curse in the Age of Energy Transition: \\ Cobalt Reserves and Conflict in Africa}

\author{Weihong Qi}

\begin{document}

\maketitle

\begin{abstract}
\linespread{1.5}\selectfont

This study reevaluates the traditional understanding of the ``political resource curse'' by examining the unique impact of energy transition metals, specifically cobalt, on local-level conflicts in Africa. Contrary to previous studies that primarily focus on high-value minerals and their political outcomes resulted from substantial economic revenues, this study investigates cobalt's influence on local conflict. Despite its strategic importance, cobalt's limited commercial value presents a unique yet critical case for analysis. Different with the prevailing view that links mineral reserves with increased conflict, this research finds that regions rich in cobalt experience a reduction in conflict. This decrease is attributed to enhanced government security measures, which are implemented independently of the economic benefits derived from cobalt as a commodity. The study utilizes a combination of georeferenced data and a difference-in-difference design to analyze the causal relationship between cobalt deposits and regional conflict. The findings suggest that the presence of cobalt deposits leads to enhanced security interventions by governments, effectively reducing the likelihood of non-governmental actors taking control of these territories. This pattern offers a new perspective on the role of energy transition metals in shaping conflict and governance, highlighting the need to reassess theoretical frameworks related to the political implications of natural resources with the ongoing energy revolution.

\end{abstract}

\section{Introduction}
Resource endowment is both a blessing and a curse in economic and political outcomes \citep{Morrison2009, PoelhekkeSteven2013, RossMichaelL2015}. Many academic studies have addressed the political implications of resource wealth under various conditions. Particularly, the famous “political resource curse” literature provides considerable evidence of the negative influence of resource endowment on different political aspects, such as democracy, the quality of institutions, and conflict \citep{RossMichaelL2015}. Prior research has explored the ramifications of diverse resources, including petroleum \citep{smith2004oil,dunning2008crude, tsui2011more, dube2013commodity}, minerals \citep{sorens2011mineral}, forestry products \citep{price2020war, harwell2011forests}, and general commodities \citep{besley2011logic, bazzi2014economic}. While there is ongoing debate within the literature regarding the specific impacts of natural resource wealth on various states and the reasons for differing outcomes among affected populations, the primary focus has been on high-value resources like petroleum and gemstones. Consequently, the proposed explanations for the effects of resource reserves primarily focus on the consequences of commodity price volatility and its subsequent influence on income, as well as tax and non-tax revenue streams.

However, over the past decade, the emergence of energy transition metals like lithium, cobalt, and nickel has posed significant challenges to the existing analytical framework of the political resource curse. These metals, crucial for lithium batteries used in alternative energy sources to fossil fuels and high-tech industries, have gained prominence. Their applications, notably in electric vehicles (EVs) and smartphones, have seen rapid expansion. The primary deposits of these resources are found in developing countries, such as Bolivia, Chile, Argentina, and the Democratic Republic of the Congo (DRC). This has led to intense competition among major powers such as the United States and China for control over these minerals, underscoring their critical role in the evolving landscape of future energy and technology industries \citep{gulley2018china}. However, despite their strategic significance, these metals possess limited commercial value. For instance, cobalt comprises only 0.037\% of the total global trade value from 2020 to 2021\citep{OEC}. This unique characteristic of energy transition metals, being strategically imperative but relatively unprofitable, presents a challenge to the conventional framework of the political resource curse. Therefore, there is a pressing need for a revised framework to analyze the impact of transition metals on political outcomes, accompanied by new empirical studies.

This study proposes that energy transition metals, such as cobalt, have a distinctive impact on conflict with high-valued minerals. It suggests that the presence of these metals motivates governments to bolster security measures, particularly around foreign-owned mineral sites, effectively reducing the frequency of local-level conflicts in resource-rich areas, independent of commodity price fluctuations. Focusing on the influence of cobalt reserves within Africa, the findings indicate that regions endowed with cobalt experience, on average, 2.05 fewer local-level conflict incidents compared to regions without cobalt. This reduction in conflicts is attributed to enhanced governmental security measures in cobalt-rich areas, which deter non-governmental actors from gaining control over these territories and push them towards areas without cobalt. Such findings underscore the distinctive role that energy transition metals play in influencing conflict and governance strategies.

To investigate the impacts of cobalt reserves, this study integrates georeferenced datasets from the U.S. Geological Survey (USGS), the Uppsala Conflict Data Program (UCDP), the Global Terrorism Database (GTD), and The Armed Conflict Location and Event Data Project (ACLED). This data is analyzed at a spatial resolution of $1^{\circ} \times 1^{\circ}$ for all African countries, spanning the years 1989 to 2019. Employing a difference-in-differences approach, the paper leverages the sudden surge in global cobalt demand in 2002 as an exogenous shock to discern the influence of cobalt reserves on conflict. The primary model compares conflict frequencies in regions with and without cobalt reserves. The study further evaluates the impact of these reserves on the likelihood of non-government entities gaining control over territories, thereby shedding light on government security responses in regions with cobalt minerals. The findings suggest that the presence of cobalt deposits can diminish conflict incidents and reduce the probability of non-government actors seizing control, implying an escalation in governmental security measures for territorial control. Additionally, the research empirically rules out several alternative explanations, such as the influence of copper reserves and international aid.

By conceptually distinguishing strategic and commercial values of minerals, the study unveils the mechanism by which energy transition metals affect local-level conflict. In contrast to the traditional narrative of the political resource curse, this research posits that energy transition metals may confer benefits for regions possessing them. It proposes that governments bolsters security measures in mineral-rich areas, while avoiding the adverse impacts typically associated with commercial revenues, such as the tendency for insurgents to seize control for funding military operations\citep{addison2002conflict, auty2001resource}. The results contrasts the findings of most of the existing literature, which mostly find positive association between mineral reserves and conflict levels. Furthermore, the research highlights the need to reevaluate the theoretical framework concerning the political implications of resources in the context of the ongoing energy revolution.

This paper also contributes to the understanding of how mining extractions impact conflict dynamics in Africa. Existing studies have documented the local and global violence in mining areas under the control of fighting groups \citep{berman2017mine}, and how resource-driven conflicts impede the economic development of African countries \citep{adhvaryu2021resources}. This research builds upon these findings by exploring the effects of an increasingly important resource in Africa. It reveals a distinct mechanism, unique compared to other mineral resources, that influences conflict on the continent.

This study is also related to the extensive literature on major power competition for resource control. Numerous research works have concentrated on the strategies employed by major powers to secure resources in developing countries. For instance, \cite{gulley2018china} examines the rivalry between the United States and China over the control of non-fuel minerals critical for emerging technologies. Further, \cite{harchaoui2021carving} proposes that official development aid is used as a means to gain privileged access to natural resources in Africa. Conversely, this research shifts the focus to the reactions of governments that possess territories rich in minerals of significant interest to these major powers.

\section{Energy Resource and Civil Conflict}

\subsection{How Does Resource Wealth Influence Conflict}
\label{sec:literature}

How do resource reserves affect levels of regional conflict? The prevailing literature focuses predominantly on high-value minerals, analyzing the impact of their extraction and revenue distribution on regional conflicts. Numerous studies explore the motivations of stakeholders in areas rich in resources, highlighting the significance of commodity price fluctuations. These fluctuations crucially influence the strategies and behaviors of regional governments, insurgents, labor forces, and the general population. The body of research, which includes both cross-national quantitative analyses and qualitative, theory-driven case studies, predominantly identifies a detrimental effect of resources on conflict levels \citep{collier2005understanding, omeje2013extractive}.

A substantial portion of the literature examines the impact of resource reserves on government structures and operations. For example, some theories propose that natural resource wealth can trigger violence by weakening government administration, reducing its capacity to prevent rebellions, or by increasing the perceived value of state control, thereby inciting new rebellions \citep{soysa2002ecoviolence, le2013fuelling}. In parallel, another segment of research analyzes the effects of mineral reserves on insurgent groups. The concept of ``contestable income,'' or the potential revenue insurgents could seize from mineral deposits, is central in this context. \cite{dube2013commodity} discusses the ``rapacity effect,'' where an increase in contestable income might escalate violence by amplifying the benefits of appropriation. This effect is attributed to resource endowment rendering the control over minerals an attractive target. Additionally, a considerable volume of research highlights the susceptibility of extractive industries, especially oil extraction sites, to armed conflict, as these locations often serve to finance military operations \citep{addison2002conflict, auty2001resource}.

In addition, a set of the literature, while not directly addressing the link between resource reserves and conflict, establishes a theoretical foundation that elucidates the impact of resources on conflict by integrating commodity prices as an intermediary factor \citep{bazzi2014econ, dal2011workers}. Specifically, this theoretical framework posits that individuals' decisions are predicated on the assessment of opportunity costs \citep{becker1968crime}. Fluctuations in commodity prices can alter the perceived costs and benefits of aligning with insurgent groups. Consequently, when the commodity originates from a labor-intensive sector, it has the potential to sway individual choices regarding affiliation with insurgents. \cite{besley2011logic} also shows that external shocks that influences the wages and aid can impact the level of violence where the institutions are weak. 

The impact of resource extraction on local communities can also escalate conflict levels. \cite{arellano2011aggravating} investigates the conflict between local communities and both mining operations and local authorities in Peru. These conflicts often stem from issues related to the distribution of benefits and labor disputes. Human rights violations and the inequitable distribution of wealth resulting from mining activities can foster grievances against the industry \citep{collier2004greed}, which may in turn exacerbate conflicts. However, much of the existing literature has focused on oil extraction sites, with limited attention to other types of resources. Recent studies indicate that the consequences of resource endowment for non-fossil fuel resources may vary significantly from those associated with fossil fuels \citep{shim2020location}.

The consensus within existing literature acknowledges a consistent association between various types of natural resources and the incidence or prolongation of civil wars \citep{lujala2005diamond, besley2011logic, RossMichaelL2015, fearon2004some}. However, the specific impacts of these resources on conflict levels demonstrate notable heterogeneity depending on their location and nature. For instance, the presence of oil deposits illustrates this variability: offshore oil reserves tend not to significantly influence a country's risk of conflict, while onshore oil deposits are often linked with considerably higher levels of conflict risk \citep{lujala2005diamond}. This distinction underscores the importance of geographical context in understanding resource-related conflicts.

\subsection{Strategic Importance vs. Commercial Values}
\label{sec:values}

The distinction between the strategic and commercial values of natural resources is crucial both conceptually and in analyzing the effect of natural resources on regional conflict. In the context of energy transition metals, the strategic significance of an energy resource could include aspects such as energy security, supply chain reliability, and their utilization in technological advancement. Conversely, commercial values are reflected in the aspects such as market liquidity, unit price, export values, and generated tax revenues. Note that the paper recognizes the interactions between the strategic and commercial aspects of natural resources. Yet, a conceptual distinction between them is critical for dissecting the impact of energy resources on conflict more effectively. The paper does not aim to exhaustively cover all aspects of the two types of values but will highlight some of the most representative elements, particularly those often cited in official documents. The theoretical section is intended to capture the essence of the concepts and pilot an innovative analytical framework.

\textbf{Strategic Values.}
\textit{Energy security} is characterized by the consistent access to energy sources at an affordable price, integrating long-term strategies to synchronize energy supply with economic expansion and environmental goals, alongside short-term measures to rectify immediate supply-demand imbalances~\citep{IEA}. Historically, petroleum has been integral to energy security, due to its extensive use in transportation, industrial output, and as a primary energy source across various economies. However, the rapid integration of lithium-ion batteries in transportation and consumer electronics underscores the growing importance of energy transition metals. \textit{Supply chain reliability} is critical for the reliable and efficient production, transport, and distribution of energy resources and components domestically. The U.S. Department of Energy has issued a comprehensive strategy to fortify America's energy supply chain, facilitating a robust transition to clean energy~\citep{USEnergy}. This strategy accentuates the vital importance of a resilient energy supply chain in combating inflation and reducing costs for American families and businesses, thereby safeguarding against the price surges caused by global supply chain disturbances. Moreover, energy transition metals such as lithium, nickel, cobalt, and rare earth elements are indispensable for \textit{technological advancement} and in determining the future geopolitical landscape among major powers. These metals, crucial for batteries, electric vehicles (EVs), and renewable energy technologies, are key to the shift towards a low-carbon economy. Considering the strategic importance of the three aspects to major powers, governments of resource-rich countries can utilize their natural endowments as bargaining chips to secure benefits for themselves. Therefore, the strategic value for these resource-endowed nations lies in their ability to leverage these assets in negotiations with major powers.

\textbf{Commercial Values.} \textit{Market liquidity}, in the context of natural resources as a type of asset, denotes the capacity to quickly buy or sell such an asset in the market without causing a significant impact on its price. This concept revolves around the ability to efficiently transform extracted resources into cash. High-value minerals, extensively analyzed in literature, typically exhibit robust market liquidity. For instance, due to petroleum's widespread utilization across various industries and economies, it can be readily converted into cash. Similarly, gold has historically functioned as a form of currency and continues to be regarded as a safe haven asset, offering a hedge against inflation and a method for diversifying investment portfolios, thereby ensuring its ease of conversion into cash. In contrast, energy transition metals, not as ubiquitously integrated into industrial applications as petroleum and requiring advanced technology for product conversion, display a lower level of market liquidity compared to petroleum and gold. This liquidity disparity further diminishes the attractiveness of energy transition metals in black markets and smuggling operations, consequently reducing their commercial value to rebel groups. A distinct but related concept is the \textit{unit price} of natural resources, which, while not solely determining their aggregated value, plays a crucial role in their transportability and tradability for economic gain. Despite their significant strategic value, energy transition metals often exhibit a relatively low unit price. For instance, one metric ton of cobalt was priced at \$63,739 in 2022~\citep{StatistaCo}, in contrast to gold, which commanded around \$64 million per metric ton~\citep{StatistaGold}, and platinum, valued at over \$33.8 million per metric ton~\citep{StatistaPlt} within the same timeframe. Consequently, the lower unit prices of energy transition metals, such as cobalt, present challenges in their transportation and trading.

While \textit{market liquidity} and \textit{unit price} represent crucial commercial values for both governments and insurgent groups, \textit{export value} and \textit{taxability} predominantly pertain to governmental revenue. Although major powers are investing in controlling energy transition metals, the limited number of countries with the requisite advanced technology to effectively exploit these resources means their total export value may not reach the levels seen with petroleum. Similarly, given that energy transition metals do not contribute as significantly to GDP percentages, they are not taxed as heavily as petroleum and precious metals. However, these metals can still yield benefits in alternative forms, such as foreign financial aid, for governments. Conversely, insurgent groups may lack access to such revenue-generating mechanisms. This disparity in benefit streams can lead to asymmetrical motivations between governments and insurgents.



\textbf{Implications.} Given the critical role of energy transition metals for global major powers, countries possessing these resources can leverage their position to negotiate benefits from major powers. Such benefits may range from individual rent-seeking advantages to targeted financial assistance for regions favored by these countries' governments. For example, as demonstrated by \cite{bommer2022home}, foreign aid is more likely to be allocated to the political leaders' native regions, influencing the distribution based on personal ties and preferences. To maintain this bargaining power, it is in the interest of these leaders and their governments to strengthen security measures surrounding energy transition mineral deposits. This ensures continued control over these valuable resources, enabling ongoing negotiations for advantageous deals with major powers.

The implications for insurgents are more complex. On one hand, rebel groups may not exhibit motivations similar to those driving the seizure of traditional energy resources such as petroleum. This discrepancy arises because energy transition metals possess strategic value but have limited commercial appeal. While governments can utilize these strategic metals to enhance their bargaining leverage with major powers, thus securing benefits in return, rebel groups are hindered by the absence of formal channels necessary for conducting equivalent exchanges. As governments intensify security measures around sites rich in transition metal minerals, and given these metals' relatively low allure, rebels may redirect their focus towards more accessible regions that are easier to dominate. On the other hand, insurgents might anticipate government strategies and deliberately target these mineral-rich areas to undermine governmental bargaining leverage with major powers, potentially escalating conflicts. I formally phrase the theoretical effects as follows:

\begin{center}
\begin{minipage}{0.9\textwidth} 
\textbf{Substitution Effect:} Insurgents shift attention to regions that are more easily to take over, in response to governments strengthening security around regions with energy transition metal minerals.
\end{minipage}
\end{center}

\begin{center}
\begin{minipage}{0.9\textwidth} 
\textbf{Signaling Effect:} Insurgents escalate their attacks on regions rich in energy transition metal minerals, foreseeing the strategic value these resources could provide the government in bargaining with major powers.
\end{minipage}
\end{center}

The prevailing effect influencing insurgent tactics remains ambiguous. Should the \textit{Substitution Effect} prevail, an increase in conflicts within regions without energy transition minerals would be anticipated. Conversely, should the \textit{Signaling Effect} take precedence, a surge in conflicts in areas rich in energy transition metals is expected.

\subsection{Cobalt: Important Resource with Limited Commercial Values}
\label{sec:cobalt}

As outlined in Section~\ref{sec:values}, cobalt is characterized as an energy resource with significant strategic importance yet comparatively lower commercial value. This strategic valuation stems from cobalt's pivotal role in advancing green energy technologies and facilitating the transition towards a low-carbon economy. The increasing emphasis on energy transition and the widespread adoption of lithium-ion batteries in high-tech sectors have amplified cobalt's relevance in technological innovation. Notably, the expanding electric vehicle (EV) market has surged the demand for cobalt, owing to its essential function in lithium-ion batteries, a key component for EVs. In 2021, the EV industry's cobalt consumption outstripped other applications, representing 34\% of the total demand. This figure is anticipated to grow, with projections suggesting EVs will account for half of the cobalt demand by 2026~\citep{CobaltIns2024}, driven by the global trend towards sustainable and green transportation solutions. Furthermore, cobalt's significance extends to the manufacture of lithium-ion batteries for a variety of high-tech industries and consumer electronics, including smartphones and laptops, where its contribution to battery energy density is vital for enhancing device performance and longevity. This escalating demand, coupled with the shift away from petroleum dependency, especially from OPEC countries, underscores cobalt's growing role in energy security and the global energy landscape.

\begin{figure}[t]
    \centering
    \includegraphics[width=\linewidth]{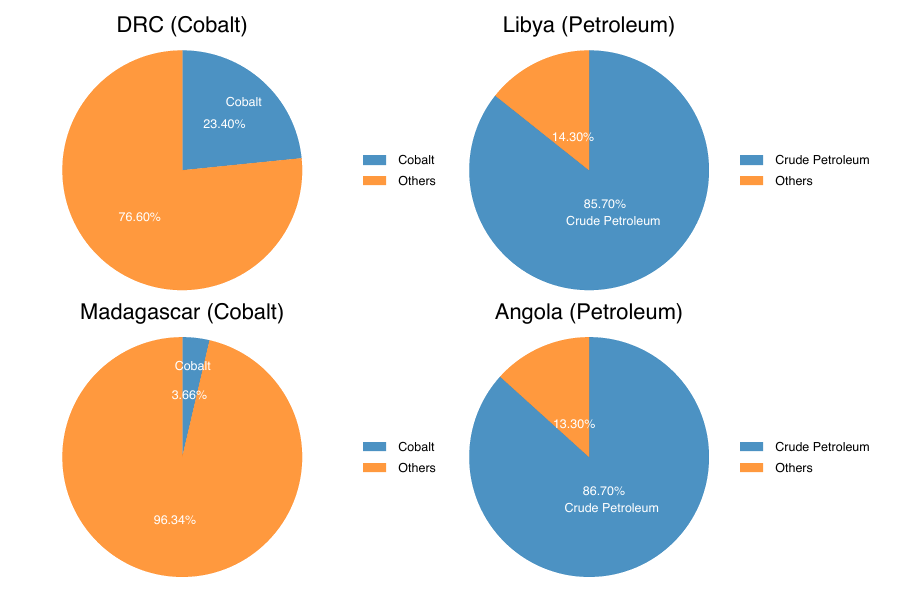}
    \caption{Percentage of Exports for Cobalt and Petroleum Among African Leading Producers}
    \label{fig:export}
\end{figure}

In contrast, the commercial value of cobalt is relatively constrained. Firstly, cobalt does not exhibit the high \textit{market liquidity} and \textit{unit prices} characteristic of other high-value resources extensively documented in the literature, such as petroleum, gold, and diamonds. Its conversion into economic benefits is not straightforward, necessitating advanced technology for its utilization and seldom being considered a safe asset. Moreover, as illustrated in Figure~\ref{fig:export}, cobalt exports from leading producers do not match the scale seen in petroleum-exporting countries. Specifically, the Democratic Republic of the Congo (DRC) and Madagascar, Africa's foremost cobalt producers, account for 23.4\% and 3.68\% of their total export values, respectively. Conversely, for leading petroleum producers like Libya and Angola, crude petroleum exports constitute over 85\% of their total export values. This disparity emerges because only a select number of countries possess the technology required to utilize cobalt effectively. Given that exports are a means of acquiring foreign currency, this suggests cobalt's limited capacity to generate foreign currency reserves compared to petroleum. Additionally, cobalt's contribution to the GDP is minimal when contrasted with other high-value resources, resulting in less substantial taxation. The disparity between cobalt's strategic importance and its commercial value stems from the fact that only a few developed countries possess the technology required to transform it into marketable products. This gap is unlikely to be closed in the short to medium term.



\subsection{Strategically Vital Resources and Security Measures}

Historically, the global distribution of strategic resources such as coal and petroleum has been markedly uneven, precipitating significant geopolitical tensions \citep{le2004geopolitical}. This strategic significance has frequently necessitated the intervention of national security forces in securing access and control. In the recent two decades, the critical role of strategic metals, including cobalt and others pivotal in energy transition and high-tech industries, is increasingly acknowledged. Nations deficient in domestic reserves of these essential materials are compelled to rely on imports. Aware of this, political elites in countries rich with strategic minerals are motivated to secure access to these resources to enhance their leverage in negotiations with major global powers, seeking varied benefits in return. This situation can be illustrated with the example when China initiated the Forum on China-Africa Cooperation in 2000. This forum laid the groundwork for China's prominent role in accessing Africa's essential energy transition minerals. It framed the Sino-African relationship as mutually beneficial: China committed to constructing critical infrastructure, including roads, dams, airports, bridges, mobile networks, and power plants throughout Africa. In return, this infrastructure development allowed China to gain access to the crucial minerals necessary to fuel its expanding economy, showing a strategic exchange where both parties purportedly benefit~\citep{kara2023cobalt}.

The governments that control these mineral resources, recognizing the potential bargaining power they offer, deliberately enhance security measures in these regions. Military and other security forces are consistently present in mining areas. The experiences documented in~\citep[p. 22]{kara2023cobalt} present both the influence of major powers of the world and the heightened security measures in the cobalt mining regions:

\begin{quote}
    As of my last ground count in November 2021, there were nineteen major industrial copper-cobalt mining complexes operating in Haut-Katanga and Lualaba Provinces, fifteen of which were owned or financed by Chinese mining companies. Most of the Chinese-owned mining sites I visited were secured either by a military force called the FARDC or the elite Republican Guard. Other industrial sites and many informal mining areas are guarded by any array of armed units, including the Congolese National Police, the mining police, private military contractors, and informal militias.
\end{quote}

Moreover, the escalation in security forces could also stem from proactive measures to prevent existing conflicts from affecting mineral operations, thereby serving as a deterrent to both rebel groups and the local community. This strategy is aimed at maintaining stable operations within these critical sectors, ensuring that disruptions do not compromise the extraction and processing of cobalt minerals. The presence of security forces, often seen as a necessary response to the volatile environment, can be interpreted as both a protective measure for the assets and a control mechanism over the local populace. Such dynamics are complex and multifaceted, as demonstrated in \cite[p. 159-160]{kara2023cobalt}, which presents another case that highlights the broader implications and motivations behind these increased security measures of the governments:

\begin{quote}
    Militia skirmishes and ethnic conflict continue to be a way of life across the Copper Belt, especially around Kolwezi. As a result, Kolwezi has the heaviest concentration of soldiers and security forces in the region. Most of the city’s major mining sites are secured by FARDC, Republican Guard, or both. In the early years of my research, the mining sites were also under the watchful eyes of the former governor of Lualaba Province and staunch ally of Joseph Kabila, Richard Muyej Mangez Mans.
\end{quote}

Observing the increased security forces near cobalt mining areas, despite cobalt's lower commercial value compared to traditional energy resources, can diminish rebel groups' motivation to seize control of these territories due to the increased difficulty in overtaking them. Consequently, rebel groups may shift their focus to other regions that are easier to control. This leads to what is termed the \textit{Substitution Effect}. However, the presence of increased security measures in mining regions might also signal the government's high valuation of these minerals, potentially enhancing the rebels' motivation to capture these areas and undermine the government's bargaining power. This tendency is the \textit{Signaling Effect}. The reaction from the rebel groups is unclear and to be investigated in the empirical section of the paper.

\section{Cobalt Mining Issues in Africa}

Cobalt production in the Democratic Republic of the Congo (DRC) is a significant aspect of the global supply chain for this critical metal, which is a crucial component in electric vehicle batteries, computers, and cell phones. The DRC is the world's largest producer of cobalt. In 2022, the DRC's cobalt mine production totaled approximately 130,000 metric tons, which was a peak production volume and accounted for nearly 70\% of worldwide cobalt mine production. The cobalt mining industry in the DRC, however, faces several challenges, including human rights issues, particularly in artisanal and small-scale mining (ASM) operations. These operations, which account for 15 to 30 percent of the Congolese cobalt production, have been subject to scrutiny due to severe human rights issues, including child labor, fatal accidents, and violent clashes. Tilwezembe, notably, is recognized as one of the most perilous industrial mining sites, notorious for its extensive use of child labor. The author of Cobalt Red highlights the grim reality of these conditions by sharing his experience: ``the final tally of my interviews was this—twelve men and boys grievously injured and seven children buried alive at Tilwezembe''~\citep[p. 154]{kara2023cobalt}.

Nevertheless, the ASM sector is crucial for the livelihoods of millions of Congolese living in poverty, making it a complex issue to resolve. An interviewee in \cite{kara2023cobalt} describes their dependence on cobalt mining for survival, highlighting the sector's significance in the local economy and the challenging balance between economic necessity and human rights concerns:

\begin{quote}
    My father died three years ago. I am the oldest son, so it was my responsibility to earn money for my family. I started digging in the fields in the south of Fungurume with a group of boys who were my friends. We dug in small pits. Some days, we found ore; some days, we did not. We were not earning much this way, so we felt we must go to the [Tenke Fungurume] concession.
\end{quote}

There have been efforts to formalize ASM cobalt mining practices, regulating mining methods and working conditions. This formalization aims to ensure that ASM cobalt is sourced responsibly, improving mine safety and child labor standards. 
However, the absence of a universally accepted definition of ``responsible ASM'' continues to present challenges for the market acceptance of ASM cobalt. These complexities are further compounded by the global demand for cobalt, which drives both formal and informal sectors toward maximizing extraction. Stakeholders, including international corporations and local governments, are under pressure to balance economic growth with ethical practices.





\section{Global Shifts to Energy Alternatives}

Because the production of energy is completely determined by the location of minerals, it raises concerns of the non-producers. The external dependence on energy producers can have a significant impact on national security, particularly if the foreign producer is a significant political or military adversary. Furthermore, the susceptible to price spikes can also increase inflation, slow economic growth, and put a strain on public finances. To mitigate these risks, many countries have sought to reduce their dependence on foreign oil by developing alternative energy sources or increasing domestic oil production, such as the development of shale gas in the United States.

In 2008, the uncertainties due to the financial crisis and the geopolitical tensions resulted in a price surge of petroleum. The high price of oil led to increased interest in alternative energy sources, including EVs. During this time, several major automakers, such as General Motors and Tesla, began to develop and launch electric vehicles. This marked a turning point for the industry and set the stage for continued growth and development in the years to come.

In addition to the innovations in commercial EVs, the governments issued policies to promote the industry and the diversity in energy consumption. The US government began subsidizing electric cars in 2010 as part of the American Recovery and Reinvestment Act (ARRA). The act provided tax credits of up to \$7,500 for individuals who purchased electric vehicles. The goal of the subsidies was to promote the production and purchase of electric cars and to encourage the development of new technologies in the field. Over time, the subsidies were modified, and some were phased out as electric vehicles became more widespread, but many incentives remain in place today. 
In 2022, the President’s Inflation Reduction Act (IRA) represents the largest investment in climate and energy in American history, aiming to ``securing America’s position as a world leader in domestic clean energy manufacturing''~\citep{IRA}. This act continues to support the clean energy sector and facilitates the ongoing transition to alternative energy sources.

Starting in 2009, the Chinese government also initiated subsidies for electric cars as part of its strategy to encourage the adoption of new energy vehicles and decrease reliance on fossil fuels. The subsidies, which have been adjusted and refined over the years, are available for both private consumers and commercial buyers, and are aimed at reducing the cost of ownership for electric cars and making them more accessible to the general public. The subsidies have been an important factor in promoting the growth of the electric car market in China and have contributed to the country's world leading position in the development and production of EVs. In the ``Made in China 2025'' national plan, the central government of China emphasizes to ``support the development of electric and fuel cell vehicles, and master core technologies for vehicle decarbonization, informatization, and intelligence''~\citep{China2025}.

The United States and China are only two examples in a broader global movement towards a shift in energy sources, supported extensively by governmental policies around the world. This trend reflects a collective international effort to transition from traditional fossil fuels to cleaner, sustainable energy alternatives like EVs. Countries across Europe, Asia, and the Americas are implementing various incentives such as subsidies, tax credits, and regulatory support to accelerate this shift. These measures are designed not only to reduce the environmental impact of energy consumption but also to secure long-term energy independence and stimulate technological innovation in the sector. As a result, the global energy landscape is witnessing a transformative shift, driven by both ecological imperatives and strategic economic considerations.



\section{Data and Measurement}
\subsection{Data}

This research aggregates data from a variety of sources, including five georeferenced datasets, data on world trade flows, commodity prices, and national-level statistics. The georeferenced data include the global distribution of selected mines, deposits, and districts of critical minerals from the U.S. Geological Survey (USGS)\footnote{https://mrdata.usgs.gov/mineral-operations/}, the Georeferenced Event Dataset (GED) from the Uppsala Conflict Data Program (UCDP)~\citep{sundberg2013introducing}, the Armed Conflict Location and Event Data Project (ACLED)~\citep{raleigh2015armed}, Global Terrorism Database (GTD)~\citep{lafree2007introducing} and the Geocoded Global Chinese Official Finance Dataset from AidData~\citep{dreher2021aid}.

The USGS dataset is utilized to pinpoint various mining deposits across Africa, including those of cobalt and copper. This dataset provides detailed location information and descriptions of mines, deposits, and mining districts for these essential minerals. The UCDP GED offers event-level data on lethal incidents, capturing the instances of organized violence. This dataset details the actors involved, conflict dyads, the conflict's nature, geographical coordinates, and the precise dates of the incidents. The GED data spans from 1989 to 2019, documenting a total of 53,872 conflict events. ACLED's georeferenced data disaggregates and examines global conflict and protest event data~\citep{raleigh2015armed}. It provides comprehensive information on the timing, actors, locations, fatalities, and types of events related to political violence and protests. This study includes ACLED data on various conflict event types, such as territorial takeovers by non-state actors. The GTD dataset provides information on global terrorist events spanning from 1970 to 2020. The Geocoded Global Chinese Official Finance Dataset identifies projects financed by the Chinese Government between 2000 and 2014, covering 3,485 projects with a total official financing amount of \$273.6 billion.

To make the paper comparable with previous political resource curse research and assess the influence of commodity prices on conflict, this study employs the IMF Commodity Price Index\footnote{https://data.imf.org/?sk=471dddf8-d8a7-499a-81ba-5b332c01f8b9} as a proxy for commodity prices under investigation. Additionally, the research incorporates national-level data, including GDP, GDP growth, and population figures from the World Bank\footnote{https://databank.worldbank.org/} and polity scores from the Polity5 dataset~\citep{marshall2020polity5} to adjust for variations across countries and over time.

\subsection{Measurement and Variables}
\textbf{Measurement.} This study integrates georeferenced cobalt mining extraction data with conflict information, employing a spatial resolution of $1^{\circ} \times 1^{\circ}$ to assess the impact of cobalt deposits on conflict incidence. Utilizing this approach, Africa is segmented into 2,806 distinct areas, enabling a more precise estimation of the effects beyond  pooled panel regressions. Figure~\ref{fig:grids} illustrates the method used to partition Africa into areas of high spatial resolution and maps the distribution of conflict events as derived from the dataset.

\textbf{Dependent Variables.} The primary dependent variable in this study, \texttt{Conflict}, denotes the count of conflict incidents as recorded in the UCDP database. The UCDP database delineates conflict as ``a contested incompatibility that concerns government and/or territory where the use of armed force between two parties, one of which must be the government of a state, leads to a minimum of 25 battle-related fatalities within a single calendar year'' \citep{gleditsch2002armed}. This research employs the frequency of conflicts as its dependent variable. Furthermore, the study incorporates the variable \texttt{Non-Government Actor Overtakes Territory} from the ACLED dataset to explore the dynamics whereby insurgents may either find it more challenging to seize territories in cobalt-rich regions due to enhanced government security measures or may escalate their attacks in anticipation of these areas' strategic significance to the government. ACLED characterizes this variable as ``A non-state actor wins control and/or subdues government forces, and/or secures territory where they can operate without opposition and are considered to have exclusive control over force within that territory'' \citep{raleigh2015armed}. In this research, the variable is treated as binary to gauge the likelihood of insurgents successfully capturing cobalt-rich territories compared to non-cobalt areas. 

\textbf{Independent Variables.} The independent variable in this study is the presence of cobalt deposits, as identified through USGS data. This data set offers precise latitude and longitude details for each mineral deposit, enabling the accurate determination of deposit locations. The binary variable \texttt{Cobalt} is employed to indicate the presence or absence of a cobalt deposit within a region. Additionally, for the purpose of empirical test for alternative mechanisms, I utilize another binary variable, \texttt{Copper}, which serves a similar function but for identifying the geographical presence of copper deposits. This approach allows for a comparative analysis of the impact of different mineral deposits on \texttt{Conflict}.

\textbf{Controls.} In the analysis, I incorporate a variety of country-level control variables, including the polity score, GDP, GDP growth rate, and population size. The polity score, which varies from -10 (representing complete autocracy) to 10 (indicating fully consolidated democracy), serves as a measure of the political regime type. Data on GDP and GDP growth rates are sourced from the World Bank dataset, providing insights into the economic scale and growth trajectory of a nation. The population variable is utilized to gauge the size of a country. Additionally, to explore alternative explanations, I include commodity prices, utilizing the IMF Commodity Price Index. This allows for a comparison with prior studies on the political resource curse and facilitates the examination of alternative mechanisms.


\section{Identification Strategy}

\subsection{Demand shock from China in 2002}

\begin{figure}[t]
    \centering
    \includegraphics[width=\linewidth]{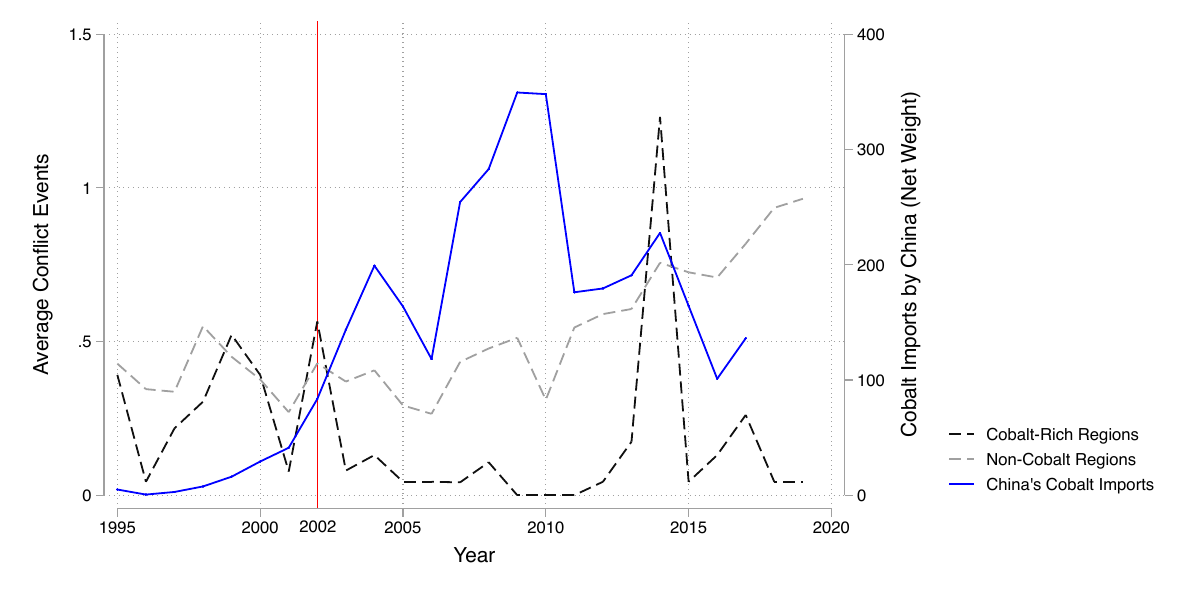}
    \caption{Cobalt Export and Conflict}
    \label{fig:did}
\end{figure}

This study adopts a difference-in-difference (DID) design with the treatment to be the sudden demand shock of cobalt because of the demand shock from China in 2002. As shown in Figure~\ref{fig:did}, the trend in conflict events follows a parallel trajectory prior to the upswing in demand and price witnessed in 2002. Subsequent to this point, a noticeable divergence between the cobalt-endowed areas and the other areas becomes evident.  Note that there is an unusual surge in conflict in year 2014, which could be due to the escalation of South Sudanese Civil War~\citep{SouthSudan} and riots initiated by Boko Haram in the year~\citep{BokoHaram}. 

The notable surge in cobalt's price and demand from China around 2002 is not a random or isolated event. This timeframe aligns with China's accession to the World Trade Organization, marking the onset of its rapid industrialization and urbanization phase. Such development required a substantial import of raw materials, with cobalt being paramount for its critical use in battery production, aerospace, electronics, and other advanced technological sectors. Although this increase predates the advent of smartphones and the expanding market for commercial electric vehicles (EVs), it represents the inception of a sustained demand for cobalt. The enduring significance of cobalt in the energy transition distinguishes it from resources that experience only a singular wave of demand and price fluctuation, highlighting it as a notable case to study the resource deposit effect on conflict.


\subsection{Empirical model specification}

To test the overall effect of cobalt deposits on conflict, I specify the empirical model as follows:

\begin{center}

$Conflict_{it} = \alpha_0 + \alpha_1 Cobalt_{i} \times Post_{2002} + \alpha_2 Cobalt_i + \alpha_3 Post_{2002} + \alpha_4 C_{it} + \alpha_5 \gamma_c + \alpha_6 \eta_t + \epsilon_{it}$ (1)

\end{center}

where $Conflict_{it}$ is the conflict incidence in unit $i$ at year $t$; $Cobalt_i$ is the binary indicating cobalt deposits in unit $i$, $Post_{2002}$ is the dummy variable to capture pre-treatment or post-treatment of the observation; $C_{it}$ is the vector of country-level controls for unit $i$ at time $t$, including GDP, GDP growth, population, and polity score; $\gamma_c$ is the country fixed effect for unit $i$; $\eta_t$ is the year fixed effect; $\epsilon_{it}$ is the error term. The coefficient of the interest is $\alpha_1$, which captures the treatment effect of of cobalt endowment in 2002.

Then, to track the source of conflict level change and test the mechanism for cobalt deposits to influence conflict, I specify the following empirical model: 

\begin{center}

$Overtake_{it} = \alpha_0 + \alpha_1 Cobalt_{i} \times Post_{2002} + \alpha_2 Cobalt_i + \alpha_3 Post_{2002} + \alpha_4 C_{it} + \alpha_5 \gamma_c + \alpha_6 \eta_t + \epsilon_{it}$ (2)

\end{center}

In Model (2), the approach for identification remains consistent with the previous model. The independent variable and controls are unchanged. However, the dependent variable in this model shifts focus to the binary of non-government actors successfully taking over areas with cobalt reserves. This change in the dependent variable offers indirect evidence regarding the dynamics of government security measures, reflecting the probability of non-government actors successfully overtaking such territories. This adjustment provides a nuanced understanding of the influence of government interventions in regions rich in critical resources like cobalt.

Further, to investigate whether the commodity prices of cobalt influence the level of conflict like high-value minerals, I specify the third model as follows:

\begin{center}

$Overtake_{it} = \alpha_0 + \alpha_1 Cobalt_{i} \times Price_{t-1} + \alpha_2 Cobalt_i + \alpha_3 Post_{2002} + \alpha_4 C_{it} + \alpha_5 \gamma_c + \alpha_6 \eta_t + \epsilon_{it}$ (3)

\end{center}

The coefficient of interest in model (3) is the coefficient $\alpha_1$ of the interaction term $Cobalt_{i} \times Price_{t-1}$, which is widely used in existing political resource curse research. This model operates under the premise that the commodity price of cobalt, an exogenous factor relative to local mineral production and conflict at a disaggregated level, is expected to exert insignificant influence on conflict. This is given cobalt's lower profitability compared to other minerals in previous studies. Therefore, Model (3) serves as a placebo test to ascertain whether the indirect evidence inferred from Model (2) is confounded by price variations.

\section{Results}

\subsection{Cobalt Reserves and Conflict}
\label{sec:main results}

Table~\hyperref[tab: cobalt_conflict]{1} displays the outcomes of the difference-in-differences analysis, examing the impact of cobalt deposits in a region on conflict levels. The results, as reported across Columns (1) to (3), vary based on the controls and fixed effects applied. These empirical results consistently suggest that the presence of cobalt deposits is associated with a reduction in conflict. Specifically, Column (3) shows a quantitative decrease in the number of conflict events by 2.045, taking into account all control variables along with year and country fixed effects. This outcome, indicating that cobalt deposits can alleviate local-level conflicts, stands in contrast to many previous studies that have investigated the effects of other minerals \citep{RossMichaelL2015}. The results suggest the validity of the \textit{Redirection Effect} as mentioned in Section~\ref{sec:values}. However, it is possible that the presence of cobalt reduces conflict by influencing other factors other than the deterrent effect of heightened security measures on insurgents. Thus, additional empirical analysis is necessary to unravel the underlying mechanisms driving this effect, which is presented in the next subsection.

To check potential systematic differences and potential confounders between the treatment and control groups in the study, the balance test results are reported in Table~\hyperref[tab: balance]{A.1}. These results indicate no significant correlation between cobalt deposits and any pre-treatment control variables in the empirical model. This suggests that the treatment and control groups were comparable in terms of their baseline characteristics, lending credence to the reliability of the identification strategy.

To validate the robustness of the findings with different fixed effects and clustering approaches, the results are presented in Table~\hyperref[tab: id_cluster]{A.2} using spatial grid fixed effects and clustering the standard errors at both the country and grid levels in the Appendix. Despite slight variations in the magnitude of the effects, the primary conclusions drawn from Table~\hyperref[tab: cobalt_conflict]{1} remain unchanged. This reinforces the validity of the initial findings, demonstrating that the observed effects are not significantly altered by the choice of clustering or fixed effects. 

\begin{table}[htbp]\centering
\def\sym#1{\ifmmode^{#1}\else\(^{#1}\)\fi}
\caption{Cobalt Reserves and Conflict}
\resizebox{0.6\linewidth}{!}{%
\begin{tabular}{l*{3}{c}}
\hline\hline
Dependent Variable: Conflict            &\multicolumn{1}{c}{(1)}&\multicolumn{1}{c}{(2)}&\multicolumn{1}{c}{(3)}\\
\hline
$Cobalt \times Post_{2002}$   &      -2.210\sym{**} &      -2.195\sym{**} &      -2.045\sym{**} \\
            &     (0.986)         &     (0.983)         &     (0.915)         \\
[1em]
Cobalt          &       2.017\sym{*}  &       1.942\sym{*}  &       1.559         \\
            &     (1.013)         &     (1.018)         &     (0.973)         \\
[1em]
$Post_{2002}$      &       0.135         &       0.249         &      -0.619         \\
            &     (0.152)         &     (0.193)         &     (1.156)         \\
[1em]
Polity      &                     &    -0.00752         &     0.00434         \\
            &                     &    (0.0143)         &    (0.0238)         \\
[1em]
Log(GDP)     &                     &      -0.126         &      -0.410         \\
            &                     &    (0.0985)         &     (0.280)         \\
[1em]
GDP Growth  &                     &     -0.0503         &     -0.0573         \\
            &                     &    (0.0388)         &    (0.0495)         \\
[1em]
Log(Population)     &                     &       0.226\sym{***}&       2.777         \\
            &                     &    (0.0815)         &     (1.661)         \\
[1em]
Country FE  &          No         &          No         &         Yes         \\
[1em]
Year FE     &          No         &          No         &         Yes         \\
\hline
\(N\)       &       87481         &       73772         &       73772         \\
\hline\hline
\multicolumn{4}{l}{\footnotesize \begin{minipage}{0.55\linewidth} \smallskip \textbf{Note:} The table shows the estimates from the difference-in-difference regression with varying controls and fixed effects. Column (1) reports the regression results without controls and fixed effect. Column (2) includes controls in the regression and column (3) includes country and year fixed effects. Robust standard errors in parentheses are clustered at country level.  \\  \sym{*} \(p<0.1\), \sym{**} \(p<0.05\), \sym{***} \(p<0.01\).   \end{minipage}}\\
\end{tabular}
}
\end{table}
\label{tab: cobalt_conflict}

\subsection{Mechanism}

Given the empirical results in Table~\hyperref[tab: cobalt_conflict]{1}, it is important to track the source of the decrease in conflict to uncover the mechanism. In this subsection, I examine if governments in regions with cobalt deposits are more effective in preventing non-state actors from taking control of these areas. The findings, detailed in Table~\hyperref[tab: mechanism]{2}, are derived using logistic regression to estimate the influence of cobalt deposits on the likelihood of non-state actors overtaking territories. The results presented in columns (1) to (3) consistently indicate that the presence of cobalt deposits in a region significantly reduces the chances of non-state actors gaining control. This pattern holds true when varying controls and fixed effects are applied, showing the robustness of the results. The results provide evidence suggesting that government-implemented security measures around strategically significant minerals effectively deter rebel groups from either attempting to or successfully seizing control of these territories.

Columns (4) to (6) of Table~\hyperref[tab: mechanism]{2} report on the relationship between cobalt commodity prices and regional conflict levels. Accounting for country-level variables and implementing both country and year fixed effects, the findings indicate that cobalt prices do not show a statistically significant correlation with regional conflict levels. This supports the hypothesis that the influence of cobalt deposits is primarily due to their strategic importance rather than being tied to price fluctuations. Furthermore, this outcome aligns with the notion that the economic benefits, often considered substantial in previous research, play a lesser role in assessing the impact of energy transition metals on conflict. To check the robustness of the estimation results for the mechanism, the estimation results using Probit and Ordinary Linear Regression (OLS) models are reported in the Table~\hyperref[tab: appendix_OLS]{A.4}. The results are robust across different model specifications.

\begin{table}[htbp]\centering
\def\sym#1{\ifmmode^{#1}\else\(^{#1}\)\fi}
\caption{Cobalt Reserve, Price and the Likelihood for Non-state Actor Overtakes Territory}
\resizebox{0.8\linewidth}{!}{%
\begin{tabular}{l*{6}{c}}
\hline\hline
Model:    &\multicolumn{6}{c}{Logit Regression} \\
            Dependent Variable:  &\multicolumn{6}{c}{Binary: Non-government actor overtakes territory} \\
                         \cmidrule(r{1em}){2-7}
             &\multicolumn{1}{c}{(1)}&\multicolumn{1}{c}{(2)}&\multicolumn{1}{c}{(3)}&\multicolumn{1}{c}{(4)}&\multicolumn{1}{c}{(5)}&\multicolumn{1}{c}{(6)}\\
\hline
$Cobalt \times Post_{2002}$ &      -2.704\sym{**} &      -2.589\sym{*}  &      -2.360\sym{*}  &                     &                     &                     \\
            &     (1.239)         &     (1.346)         &     (1.375)         &                     &                     &                     \\
[1em]
$Cobalt \times Price_{t-1}$ &                     &                     &                     &     -0.0125\sym{**} &    -0.00965\sym{*}  &     -0.0111         \\
            &                     &                     &                     &   (0.00518)         &   (0.00496)         &   (0.00698)         \\
Cobalt          &       0.391         &       0.335         &      -0.567         &       0.646         &       0.380         &      -0.303         \\
            &     (0.654)         &     (0.743)         &     (1.334)         &     (0.631)         &     (0.665)         &     (1.029)         \\
[1em]
$Post_{2002}$      &      -0.354         &       0.481\sym{*}  &      0.0563         &                     &                     &                     \\
            &     (0.314)         &     (0.285)         &     (1.504)         &                     &                     &                     \\
[1em]
$Price_{t-1}$  &                     &                     &                     &    -0.00199         &     0.00219         &     0.00103         \\
            &                     &                     &                     &   (0.00143)         &   (0.00143)         &    (0.0126)         \\
Controls  &          No         &          Yes         &         Yes     &    No         &          Yes         &         Yes    \\
[1em]
Country FE  &          No         &          No         &         Yes         &          No         &          No         &         Yes         \\
[1em]
Year FE     &          No         &          No         &         Yes         &          No         &          No         &         Yes         \\
\hline
\(N\)       &       65209         &       54398         &       45191         &       65209         &       54398         &       45191         \\
\hline\hline
\multicolumn{7}{l}{\footnotesize \begin{minipage}{0.75\linewidth} \smallskip \textbf{Note:} 
The table presents the results of a difference-in-differences regression using a logistic regression model. The dependent variable is a binary indicator representing whether a non-government actor has overtaken territory. Columns (1) through (3) show the effects of cobalt reserves, while Columns (4) through (6) focus on the effect of fluctuations in cobalt prices. Robust standard errors in parentheses are clustered at country level. \\
\sym{*} \(p<0.1\), \sym{**} \(p<0.05\), \sym{***} \(p<0.01\).   \end{minipage}}\\
\end{tabular}
}
\end{table}
\label{tab: mechanism}

\subsection{Robustness}

In the Appendix, I present additional results to validate the robustness of the estimation outcomes. As shown in the Figure~\ref{fig:did}, the year 2014 experienced heightened conflicts due to escalations in South Sudan and the activities of Boko Haram, marking it as a potential outlier in the dataset. To confirm the robustness of our findings, I exclude the year 2014 and re-conduct the regressions in Table~\hyperref[tab: cobalt_conflict]{1}. The revised outcomes, reported in Table~\hyperref[tab: appendix_outlier]{A.5}, affirm that the primary conclusions of the study remain consistent, even after removing the outlier year 2014.


Given that DRC is the primary producer of cobalt of the world taking over 70\% of the cobalt production, a valid concern arises regarding whether the effects estimated pertain specifically to the DRC or are representative of the entire African continent. To address this issue, I re-conduct the regressions excluding the DRC and report the outcomes in Table~\hyperref[tab: appendix_noDRC]{A.6}. The results clarify that the effects discussed in Section~\ref{sec:main results} reflect the situation across the whole African continent, not just within the DRC. Although the magnitude of the coefficients is slightly larger than those presented in Table~\hyperref[tab: cobalt_conflict]{1}, the core conclusions of the study remain unchanged.

Additionally, I utilize data from the GTD to further validate the proposed mechanism. Beyond conflicts stemming from rebel groups and local community riots, terrorist attacks represent another significant source of conflict. An increase in security measures, aimed at deterring insurgents, would logically extend to deter terrorists as well, making it more challenging for them to execute attacks. The findings presented in Table~\hyperref[tab: terrorist]{A.7} indicate that the presence of cobalt deposits correlates with a reduction in the number of terrorist attacks in the region. This supports the hypothesis that an enhanced security presence by governments in areas with valuable resources can deter potential agitators, thereby lowering the overall level of conflict.

\subsection{Alternative Mechanisms}
Several factors might confound the influence of cobalt deposits on security measures and regional conflict. The coexistence of cobalt and copper is a significant factor. Cobalt often occurs in conjunction with copper ores, particularly in the Democratic Republic of the Congo, which holds a substantial portion of the world's cobalt and copper reserves. This geological pairing means that areas rich in cobalt are often also abundant in copper, complicating the analysis of cobalt's specific impact on regional security and conflict. To ascertain the robustness of the identification strategy employed, the study employs copper deposits as a placebo independent variable. The findings are presented in Table~\hyperref[tab: copper]{3}, which utilizes Model (2) from Section 4 but substitutes the cobalt deposit variable with that of copper deposits. Upon accounting for the relevant controls and fixed effects, it is observed that copper deposits do not exhibit a statistically significant impact on regional conflict levels. This test further validates the conclusions in Section~\ref{sec:main results}.

\begin{table}[htbp]\centering
\def\sym#1{\ifmmode^{#1}\else\(^{#1}\)\fi}
\caption{Copper Reserve and the Likelihood for Non-state Actor Overtakes Territory}
\resizebox{0.9\linewidth}{!}{%
\begin{tabular}{l*{3}{>{\centering\arraybackslash}p{0.2\linewidth}}}
\hline\hline
Model:    &\multicolumn{3}{c}{Logit Regression} \\
            Dependent Variable:  &\multicolumn{3}{c}{Binary: Non-government actor overtakes territory} \\
            \cmidrule(r{1em}){2-4}
            &\multicolumn{1}{c}{(1)}&\multicolumn{1}{c}{(2)}&\multicolumn{1}{c}{(3)}\\
\hline
$Copper \times Post_{2002}$   &      -0.946         &      -1.228\sym{*}  &      -1.032         \\
            &     (0.599)         &     (0.694)         &     (0.812)         \\
[1em]
Copper          &      0.0230         &       0.140         &      -0.375         \\
            &     (0.407)         &     (0.407)         &     (1.070)         \\
[1em]
$Post_{2002}$      &      -0.342         &       0.496\sym{*}  &      0.0479         \\
            &     (0.316)         &     (0.284)         &     (1.506)         \\
[1em]
Controls  &          No         &          Yes         &         Yes     \\
[1em]
Country FE  &          No         &          No         &         Yes         \\
[1em]
Year FE     &          No         &          No         &         Yes         \\
\hline
\(N\)       &       65209         &       54398         &       45191         \\
\hline\hline
\multicolumn{4}{l}{\footnotesize \begin{minipage}{0.9\linewidth} \smallskip \textbf{Note:} The table shows the results from the logistic regression model that uses the copper reserve indicator as the independent variable. Column (1) presents the regression results without any controls or fixed effects. Column (2) incorporates controls into the regression, and Column (3) adds both country and year fixed effects.  Robust standard errors in parentheses are clustered at country level. \\
\sym{*} \(p<0.1\), \sym{**} \(p<0.05\), \sym{***} \(p<0.01\).   \end{minipage}}\\
\end{tabular}
}
\end{table}
\label{tab: copper}



In addition, foreign financial aid often plays a pivotal role in influencing access to mineral resources and the conflict in developing countries~\citep{nielsen2011foreign}. This aid can come in various forms, including development assistance, infrastructural investments, and economic grants. Such aid is frequently directed towards sectors that are critical for the extraction and processing of minerals, like transportation, energy, and industrial development. This influx of foreign aid and investment is often accompanied by complex geopolitical interests. Donor countries or organizations may seek to secure favorable terms for access to these minerals, which are crucial for various industries, including electronics and renewable energy. Existing studies also indicate the increasing of official foreign aid could reduce the level of conflict~\citep{gutting2017donor}. To assess the impact of foreign aid on regions with cobalt deposits, this study utilizes data from AidData to track the flow of aid from China to cobalt-rich areas. The findings are detailed in Table~\hyperref[tab: aid]{4}. The analysis reveals a positive but not statistically significant correlation between cobalt deposits and the receipt of official financial aid from China. This suggests the official aid from China is not a significant factor that may influence the chance for non-state actors to control the resource-rich regions.

\begin{table}\centering
\def\sym#1{\ifmmode^{#1}\else\(^{#1}\)\fi}
\caption{Cobalt Deposits and Financial Aid from China }
\resizebox{0.6\linewidth}{!}{%
\begin{tabular}{l*{3}{c}}
\hline\hline
            Dependent Variable:  &\multicolumn{3}{c}{Official Financial Aid from China} \\
            &\multicolumn{1}{c}{(1)}&\multicolumn{1}{c}{(2)}&\multicolumn{1}{c}{(3)}\\
\hline
$Cobalt \times Post_{2002}$   &       4.258         &       3.434         &       4.384         \\
            &     (7.225)         &     (7.299)         &     (6.931)         \\
[1em]
Cobalt          &       0.859         &       0.803         &      -1.476         \\
            &     (0.974)         &     (0.983)         &     (2.314)         \\
[1em]
$Post_{2002}$      &       5.304\sym{***}&       5.374\sym{***}&       8.302         \\
            &     (1.311)         &     (1.575)         &     (9.577)         \\
[1em]
Controls  &          No         &          Yes         &         Yes     \\
[1em]
Country FE  &          No         &          No         &         Yes         \\
[1em]
Year FE     &          No         &          No         &         Yes         \\
\hline
\(N\)       &       42952         &       37586         &       37586         \\
\hline\hline
\multicolumn{4}{l}{\footnotesize \begin{minipage}{0.5\linewidth} \smallskip \textbf{Note:} 
The table presents estimates from a linear regression , with the official financial flow from China as the dependent variable. Column (1) shows the regression results without controls or fixed effects. Column (2) adds controls into the regression model, and Column (3) further incorporates both country and year fixed effects to refine the analysis. Robust standard errors in parentheses are clustered at country level.\\
\sym{*} \(p<0.1\), \sym{**} \(p<0.05\), \sym{***} \(p<0.01\).   \end{minipage}}\\
\end{tabular}
}
\end{table}
\label{tab: aid}

\section{Conclusions and Discussions}


The conventional political resource curse analytical framework mainly focuses on high-value minerals, regarding the impact of global commodity price fluctuations as external shocks and thus examining how resource deposits influence various political outcomes. However, cobalt differs from previously studied resources in that its commercial value is relatively modest compared to its strategic importance. This disparity is unlikely to be bridged in the short to medium term due to the advanced technologies required to process cobalt into lithium-ion batteries, which are only available in a select number of developed countries. This characteristic—significant strategic value combined with limited commercial value—renders cobalt a unique and crucial case for study in the age of energy transition. 

This paper makes a conceptual distinction between the strategic and commercial values of natural resources, aiming to develop an innovative analytical framework within the context of energy transition. Strategic values, which include aspects such as energy security, supply chain reliability, and technological advancement, are frequently highlighted in governmental documents. These values drive major powers to compete for control over these minerals and encourage governments of cobalt-rich countries to secure access to these resources in exchange for benefits from foreign governments. On the other hand, commercial values, such as market liquidity, unit price, and taxability, are directly associated with the economic returns derived from these minerals. Cobalt, which has high strategic value but relatively low commercial value, prompts governments to enhance security measures near mining sites. This response aims to protect these valuable assets from potential threats, including insurgent groups. However, the impact on insurgents is not straightforward due to a complex trade-off: while increased security presents a significant barrier, the strategic importance of disrupting such sites could potentially offer high rewards.

Utilizing georeferenced datasets and a difference-in-differences design, this research identifies the causal impact of cobalt reserves on local conflict dynamics. The empirical results indicate that increased security measures implemented by governments, aimed at securing access to mining sites and protecting foreign-owned properties, effectively deter insurgent groups from cobalt-rich regions, consequently leading to a reduction in the level of conflict in those areas. These findings stand in contrast to much of the existing literature, which typically reports a positive correlation between mineral reserves and conflict intensity. Additionally, the research effectively rules out alternative mechanisms, such as the influence of copper reserves and foreign financial aid, thereby strengthening the case that the observed effects are specifically related to cobalt. 

This research has several limitations that should be noted. First, it lacks georeferenced data for military deployments, resulting in an absence of direct measurements of security forces. Second, due to data availability constraints, it remains unclear how foreign ownership—particularly the increased Chinese investment in mining sites—affects the level and quality of security measures implemented near the minerals. Future studies should aim to directly measure these variables and gather concrete evidence to further explore this topic. Additionally, cobalt is not the only energy transition metal, nor is conflict the sole political outcome associated with the global energy transition. Future research could benefit from applying the conceptual distinction between strategic and commercial values to other energy metals and examining their broader political and economic impacts during the age of global energy transition.


\clearpage
\bibliographystyle{apsr}
\bibliography{cobalt}

\clearpage
\appendix
\section{Appendix}

\renewcommand{\thetable}{\thesection.\arabic{table}} 
\setcounter{table}{0}
\renewcommand{\thesection}{\Alph{section}} 

\renewcommand{\thefigure}{\thesection.\arabic{figure}}

\setcounter{figure}{0}

\begin{table}[h]
\centering
\def\sym#1{\ifmmode^{#1}\else\(^{#1}\)\fi}
\caption{Summary Statistics}
\begin{tabular}{l*{1}{ccccc}}
\hline\hline
                    &       Obs. &        Mean &      Std. Dev. &        Min &         Max \\
\hline
Conflict            &       88083&        0.52&    5.795058&        0.00&     1089.00\\
Non-government Actors Overtakes Territory                  &             65736&        0.02&    .1218541&        0.00&        1.00 \\
Cobalt                  &       88114&        0.01&    .1213397&        0.00&        1.00\\
Copper                  &       88114&        0.05&    .2196323&        0.00&        1.00\\
Polity Score              &       75967&        0.41&    5.003525&      -10.00&       10.00\\
GDP                 &       78320&    49834.50&    86032.71&       75.95&   574183.83\\
GDP Growth          &       78303&        3.74&    5.517546&      -50.34&      149.97\\
Population          &       81405&       30.08&    30.68645&        0.07&      203.30\\
\hline\hline
\multicolumn{6}{l}
{\footnotesize \begin{minipage}{0.9\linewidth} \smallskip \textbf{Note:} 
GDP is measured in millions of dollars, and GDP growth is measured in the number of percentage. Population is measured in millions. \end{minipage}}
\end{tabular}
\end{table}
\label{tab: summary}

\begin{table}[htbp]\centering
\def\sym#1{\ifmmode^{#1}\else\(^{#1}\)\fi}
\caption{Balance Test}
\resizebox{0.7\linewidth}{!}{%
\begin{tabular}{l*{4}{c}}
\hline\hline
            &\multicolumn{1}{c}{(1)}&\multicolumn{1}{c}{(2)}&\multicolumn{1}{c}{(3)}&\multicolumn{1}{c}{(4)}\\
Dependent Variable:            &\multicolumn{1}{c}{GDP Growth}&\multicolumn{1}{c}{Log(GDP)}&\multicolumn{1}{c}{Log(Population)}&\multicolumn{1}{c}{Polity}\\
\hline
Cobalt          &      0.0353         &     0.00154         &   0.0000358         &     0.00917         \\
            &    (0.0291)         &   (0.00365)         &  (0.000528)         &    (0.0324)         \\
[1em]
GDP Growth  &                     &   -0.000678         &    0.000996         &      0.0293         \\
            &                     &   (0.00370)         &  (0.000885)         &    (0.0248)         \\
[1em]
Log(GDP)     &      -0.232         &                     &      0.0870\sym{***}&       0.836         \\
            &     (1.260)         &                     &    (0.0208)         &     (0.775)         \\
[1em]
Log(Population)     &       7.758         &       1.983\sym{***}&                     &      -4.672         \\
            &     (6.053)         &     (0.442)         &                     &     (3.684)         \\
[1em]
Polity      &       0.118         &     0.00982         &    -0.00241         &                     \\
            &    (0.0966)         &   (0.00883)         &   (0.00180)         &                     \\
[1em]
Country FE  &         Yes         &         Yes         &         Yes         &         Yes         \\
[1em]
Year FE     &         Yes         &         Yes         &         Yes         &         Yes         \\
\hline
\(N\)       &       73791         &       73791         &       73791         &       73791         \\
\hline\hline
\multicolumn{5}{l}{\footnotesize \begin{minipage}{0.7\linewidth} \smallskip \textbf{Note:}   \sym{*} \(p<0.1\), \sym{**} \(p<0.05\), \sym{***} \(p<0.01\).  \end{minipage}}\\
\end{tabular}
}
\end{table}
\label{tab: balance}

\begin{table}[htbp]\centering
\def\sym#1{\ifmmode^{#1}\else\(^{#1}\)\fi}
\caption{Cobalt Reserves and Conflict (Varying Fixed Effects and Clusters)}
\resizebox{0.6\linewidth}{!}{%
\begin{tabular}{l*{3}{c}}
\hline\hline
Dependent Variable: Conflict            
&\multicolumn{1}{c}{(1)}&\multicolumn{1}{c}{(2)}&\multicolumn{1}{c}{(3)}\\
\hline
$Cobalt \times Post_{2002}$ &      -2.186\sym{**} &      -2.045\sym{*}  &      -2.186\sym{*}  \\
            &     (0.990)         &     (1.073)         &     (1.175)         \\
[1em]           
Cobalt          &      -7.950         &       1.559         &      -7.950\sym{***}\\
            &     (5.708)         &     (1.165)         &     (2.735)         \\
[1em]
$Post_{2002}$      &      -0.538         &      -0.619         &      -0.538         \\
            &     (1.190)         &     (0.534)         &     (0.547)         \\
[1em]
Controls  &          Yes         &          Yes         &         Yes        \\
[1em]
\hline
Year FE     &         \ding{52}         &         \ding{52}        &       \ding{52}        \\
[1em]
Grid FE     &         \ding{52}         &                  &        \ding{52}         \\
[1em]
Country FE  &               &         \ding{52}         &               \\
[1em]
Country Cluster & \ding{52} &    &  \\
[1em]
Grid Cluster &       & \ding{52} & \ding{52} \\
\hline
\(N\)       &       73772         &       73772         &       73772         \\
\hline\hline
\multicolumn{4}{l}{\footnotesize \begin{minipage}{0.45\linewidth} \smallskip \textbf{Note:} \sym{*} \(p<0.1\), \sym{**} \(p<0.05\), \sym{***} \(p<0.01\).   \end{minipage}}\\
\end{tabular}
}
\end{table}
\label{tab: id_cluster}

\begin{table}[htbp]\centering
\def\sym#1{\ifmmode^{#1}\else\(^{#1}\)\fi}
\caption{Cobalt Reserve, Price and the Likelihood for Non-state Actor Overtakes Territory (Linear Regression)}
\resizebox{\linewidth}{!}{%
\begin{tabular}{l*{6}{c}}
\hline\hline
Model:    &\multicolumn{6}{c}{Linear Regression} \\
            Dependent Variable:  &\multicolumn{6}{c}{Binary: Non-government actor overtakes territory} \\
                         \cmidrule(r{1em}){2-7}
            &\multicolumn{1}{c}{(1)}&\multicolumn{1}{c}{(2)}&\multicolumn{1}{c}{(3)}&\multicolumn{1}{c}{(4)}&\multicolumn{1}{c}{(5)}&\multicolumn{1}{c}{(6)}\\
\hline
$Cobalt \times Post_{2002}$ &     -0.0216\sym{**} &     -0.0194\sym{**} &     -0.0164\sym{*}  &                     &                     &                     \\
            &   (0.00909)         &   (0.00862)         &   (0.00846)         &                     &                     &                     \\
[1em]
$Cobalt \times Price_{t-1}$ &                     &                     &                     &  -0.0000492         &  -0.0000463         &  -0.0000342         \\
            &                     &                     &                     & (0.0000586)         & (0.0000631)         & (0.0000619)         \\
[1em]
Cobalt          &     0.00923         &     0.00978         &    0.000279         &   -0.000537         &     0.00111         &    -0.00761         \\
            &   (0.00786)         &   (0.00739)         &   (0.00731)         &   (0.00853)         &   (0.00860)         &   (0.00848)         \\
[1em]
$Post_{2002}$      &    -0.00573\sym{***}&     0.00348\sym{***}&      0.0200\sym{***}&                     &                     &                     \\
            &   (0.00110)         &   (0.00122)         &   (0.00644)         &                     &                     &                     \\
[1em]

$Price_{t-1}$   &                     &                     &                     &  -0.0000280\sym{***}&  0.00000678         &    0.000163\sym{***}\\
            &                     &                     &                     &(0.00000724)         &(0.00000873)         & (0.0000540)         \\
[1em]
Controls  &          No         &          Yes         &         Yes     &    No         &          Yes         &         Yes    \\
Country FE  &          No         &          No         &         Yes         &          No         &          No         &         Yes         \\
[1em]
Year FE     &          No         &          No         &         Yes         &          No         &          No         &         Yes         \\
\hline
\(N\)       &       65667         &       54398         &       54398         &       65667         &       54398         &       54398         \\
\hline\hline
\multicolumn{7}{l}{\footnotesize \begin{minipage}{\linewidth} \smallskip \textbf{Note:} \sym{*} \(p<0.1\), \sym{**} \(p<0.05\), \sym{***} \(p<0.01\).   \end{minipage}}\\
\end{tabular}
}
\end{table}
\label{tab: appendix_OLS}

\begin{table}[htbp]\centering
\def\sym#1{\ifmmode^{#1}\else\(^{#1}\)\fi}
\caption{The Results without Year of 2014}
\resizebox{0.6\linewidth}{!}{%
\begin{tabular}{l*{3}{c}}
\hline\hline
Dependent Variable: Conflict            &\multicolumn{1}{c}{(1)}&\multicolumn{1}{c}{(2)}&\multicolumn{1}{c}{(3)}\\

\hline
$Cobalt \times Post_{2002}$ &      -2.224\sym{**} &      -2.163\sym{**} &      -2.008\sym{**} \\
            &     (0.979)         &     (0.978)         &     (0.911)         \\
[1em]
Cobalt          &       2.017\sym{*}  &       1.939\sym{*}  &       1.533         \\
            &     (1.013)         &     (1.016)         &     (0.977)         \\
[1em]
$Post_{2002}$      &       0.124         &       0.215         &      -0.742         \\
            &     (0.148)         &     (0.186)         &     (1.125)         \\
[1em]
Controls  &          No         &          Yes         &         Yes        \\
[1em]
Country FE  &          No         &          No         &         Yes         \\
[1em]
Year FE     &          No         &          No         &         Yes         \\
\hline
\(N\)       &       84619         &       71303         &       71303         \\
\hline\hline
\multicolumn{4}{l}{\footnotesize \begin{minipage}{0.55\linewidth} \smallskip \textbf{Note:} \sym{*} \(p<0.1\), \sym{**} \(p<0.05\), \sym{***} \(p<0.01\).   \end{minipage}}\\
\end{tabular}
}
\end{table}
\label{tab: appendix_outlier}

\begin{table}[htbp]\centering
\def\sym#1{\ifmmode^{#1}\else\(^{#1}\)\fi}
\caption{The Results without DRC}
\resizebox{0.6\linewidth}{!}{%
\begin{tabular}{l*{3}{c}}
\hline\hline
Dependent variable: Conflict            &\multicolumn{1}{c}{(1)}&\multicolumn{1}{c}{(2)}&\multicolumn{1}{c}{(3)}\\

\hline
$Cobalt \times Post_{2002}$ &      -2.355\sym{**} &      -2.385\sym{**} &      -2.219\sym{**} \\
            &     (1.146)         &     (1.132)         &     (1.049)         \\
[1em]
Cobalt          &       2.224\sym{*}  &       2.197\sym{*}  &       1.719         \\
            &     (1.164)         &     (1.157)         &     (1.122)         \\
[1em]
$Post_{2002}$      &       0.119         &       0.233         &      -0.484         \\
            &     (0.161)         &     (0.195)         &     (1.157)         \\
[1em]
Controls  &          No         &          Yes         &         Yes        \\
[1em]
Country FE  &          No         &          No         &         Yes         \\
[1em]
Year FE     &          No         &          No         &         Yes         \\
\hline
\(N\)       &       81742         &       68217         &       68217         \\
\hline\hline
\multicolumn{4}{l}{\footnotesize \begin{minipage}{0.45\linewidth} \smallskip \textbf{Note:}  \sym{*} \(p<0.1\), \sym{**} \(p<0.05\), \sym{***} \(p<0.01\).   \end{minipage}}\\
\end{tabular}
}
\end{table}
\label{tab: appendix_noDRC}

\begin{table}[htbp]\centering
\def\sym#1{\ifmmode^{#1}\else\(^{#1}\)\fi}
\caption{Cobalt Reserve and the Likelihood of Terrorist Attacks}
\resizebox{0.6\linewidth}{!}{%
\begin{tabular}{l*{3}{c}}
\hline\hline
Model:    &\multicolumn{3}{c}{Logit Regression} \\
            Dependent Variable:  &\multicolumn{3}{c}{Binary: Terrorist Attack} \\
                         \cmidrule(r{1em}){2-4}
            &\multicolumn{1}{c}{(1)}&\multicolumn{1}{c}{(2)}&\multicolumn{1}{c}{(3)}\\
\hline
$Cobalt \times Post_{2002}$ &      -0.178         &     -0.1000         &      -0.729\sym{**} \\
            &     (0.249)         &     (0.328)         &     (0.323)         \\
[1em]
Cobalt          &       0.743\sym{***}&       0.531\sym{***}&       1.110\sym{***}\\
            &     (0.215)         &     (0.189)         &     (0.360)         \\
[1em]
$Post_{2002}$      &      -0.392\sym{**} &      -0.480\sym{***}&      -0.478         \\
            &     (0.190)         &     (0.149)         &     (1.611)         \\
[1em]
Controls  &          No         &          Yes         &         Yes        \\
[1em]
Country FE  &          No         &          No         &         Yes         \\
[1em]
Year FE     &          No         &          No         &         Yes         \\
\hline
\(N\)       &       87512         &       73791         &       70775         \\
\hline\hline
\multicolumn{4}{l}{\footnotesize \begin{minipage}{0.5\linewidth} \smallskip \textbf{Note:}  \sym{*} \(p<0.1\), \sym{**} \(p<0.05\), \sym{***} \(p<0.01\).   \end{minipage}}\\
\end{tabular}
}
\end{table}
\label{tab: terrorist}

\begin{figure}[t]
    \centering
    \includegraphics[width= 0.8\linewidth]{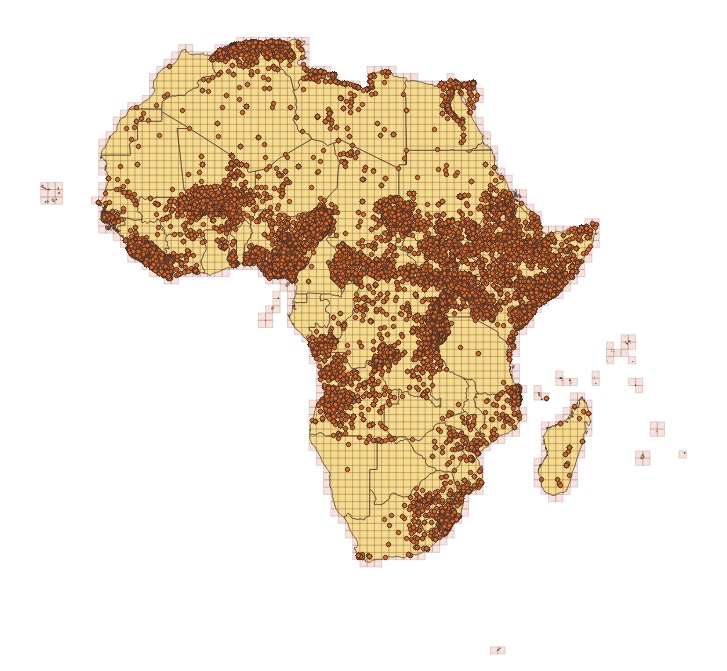}
    \caption{Geographical Segmentation and Conflict Distribution in Africa (1989 - 2019)}
    \label{fig:grids}
\end{figure}

\end{document}